\documentclass[a4paper]{article}
\usepackage{hyperref}

\usepackage{multirow}
\usepackage{tabulary}
\usepackage{setspace}

\usepackage{INTERSPEECH2022}

\title{Improving Mispronunciation Detection with Wav2vec2-based Momentum Pseudo-Labeling for Accentedness and Intelligibility Assessment}
\name{Mu Yang$^1$, Kevin Hirschi$^2$, Stephen D. Looney$^3$, Okim Kang$^2$, John H. L. Hansen$^1$}
\address{
  $^1$Center for Robust Speech Systems (CRSS), University of Texas at Dallas, Richardson, TX, USA\\
  $^2$Northern Arizona University, Flagstaff, AZ, USA\\
  $^3$Pennsylvania State University, State College, PA, USA}
\email{$^1$\{mu.yang, john.hansen\}@utdallas.edu, \\$^2$\{kevin.hirschi, okim.kang\}@nau.edu, $^3$sdl16@psu.edu}

\begin{document}
\maketitle
\begin{abstract}
Current leading mispronunciation detection and diagnosis (MDD) systems achieve promising performance via end-to-end phoneme recognition. One challenge of such end-to-end solutions is the scarcity of human-annotated phonemes on natural L2 speech. In this work, we leverage unlabeled L2 speech via a pseudo-labeling (PL) procedure and extend the fine-tuning approach based on pre-trained self-supervised learning (SSL) models. Specifically, we use Wav2vec 2.0 as our SSL model, and fine-tune it using original labeled L2 speech samples plus the created pseudo-labeled L2 speech samples. Our pseudo labels are dynamic and are produced by an ensemble of the online model on-the-fly, which ensures that our model is robust to pseudo label noise. We show that fine-tuning with pseudo labels achieves a 5.35\% phoneme error rate reduction and  2.48\% MDD F1 score improvement over a labeled-samples-only fine-tuning baseline. The proposed PL method is also shown to outperform conventional offline PL methods. Compared to the state-of-the-art MDD systems, our MDD solution produces a more accurate and consistent phonetic error diagnosis. In addition, we conduct an open test on a separate UTD-4Accents dataset, where our system recognition outputs show a strong correlation with human perception, based on accentedness and intelligibility. 
\end{abstract}

\section{Introduction}


Second-language (L2) English learners typically present accents and mispronunciations, which highly impact their intelligibility in practical communication. In recent years, Computer Aided Pronunciation Training (CAPT) tools have been developed to provide diagnosis and feedback on phonetic-level errors (phoneme substitution, deletion, insertion \cite{sudhakara2019improved, li2016improving, wu21h_interspeech, peng21e_interspeech, feng2020sed}) and prosodic-level errors (e.g. lexical stress, intonation \cite{korzekwa21_interspeech}). In this study, we focus on detecting phonetic-level pronunciation errors for L2 speech intelligibility and accentedness assessment.

Currently, most phonetic-level mispronunciation detection and diagnosis (MDD) systems perform end-to-end phoneme recognition on L2 speech, based on deep neural network (DNN) architectures \cite{wu21h_interspeech, peng21e_interspeech, feng2020sed, lo20b_interspeech, yan20_interspeech}. One of the challenges of training such DNNs is data sparsity, due to the laborious process of annotating perceived phonemes on L2 speech. To address this issue, the ``pre-training + fine-tuning" scheme has been shown to be effective \cite{devlin2019bert}: in pre-training stage a model was trained on external large scale unlabeled data using self-supervised learning (SSL) objectives, and then fine-tuned on the data from the downstream task with task-specific supervision. In speech realm, multiple SSL pre-trained models have been proposed \cite{baevski2020wav2vec, hsu2021hubert, wang2021unispeech, chen2021wavlm}, and have shown promising results on many downstream tasks \cite{lin2021fragmentvc, fan2020exploring, chen2021exploring}, including MDD task \cite{peng21e_interspeech}.

However, since data of the target task is limited, the vanilla fine-tuning approach may not be ideal due to domain mismatch between the pre-training and the target task \cite{gururangan2020don, dery2021should}. One solution is to use unlabeled data from the target domain \cite{chen2021exploring, gururangan2020don}. On MDD task, how to leverage unlabeled L2 speech remains unexplored. We approach this problem from a semi-supervised learning perspective based on pseudo-labeling (PL). We use Wav2vec 2.0 \cite{baevski2020wav2vec} as the SSL pre-trained model and extend the ``pre-training + fine-tuning" scheme with one additional fine-tuning stage where pseudo-labeled L2 utterances are included in training. We propose to employ a recent momentum pseudo-labeling (MPL) method \cite{higuchi2021momentum}. Unlike the conventional PL methods \cite{kahn2020self, xu20b_interspeech, park20d_interspeech}, MPL generates pseudo labels in a dynamic and online manner via teacher-student training: the online student model is trained using pseudo labels generated by an offline teacher model. The teacher model maintains a momentum-based moving average of the weights of the online model, which can be seen as an ensemble of the student model. This makes the online model robust to pseudo label noise and stabilizes training on unlabeled samples. We show that fine-tuning with MPL improves a vanilla fine-tuning baseline by 5.35\% in phoneme error rate (PER), and 2.48\% in MDD F1 score.

In addition, we take one step forward towards using the MDD model for automatic L2 speech intelligibility and accentedness assessment. We conduct an open test of our MDD model on a separate Indian-accented L2 English corpus. Through a human listening test, we show that the phoneme recognition performance of the MDD model has strong correlations with human ratings of L2 speech intelligibility and accentedness. This finding reveals the alignment between the MDD model prediction and human perception. \footnote{We provide an audio demo at \url{https://mu-y.github.io/speech_samples/mdd_IS22/}. Code will be available at \url{https://github.com/Mu-Y/mpl-mdd}.}


\section{Related work}

\textbf{MDD.} Goodness-of-pronunciation (GOP) is among the first DNN-based methods to MDD, which relies on phone posterior outputs from an automatic speech recognizer (ASR) \cite{sudhakara2019improved, li2016improving, hu2015improved} to evaluate phonetic errors. More recently, end-to-end phoneme recognition has been studied \cite{wu21h_interspeech, peng21e_interspeech, feng2020sed, yan20_interspeech, xu21k_interspeech}, among which \cite{peng21e_interspeech} and \cite{xu21k_interspeech} also explored fine-tuning Wav2vec 2.0. Our proposed method differs from them in that we investigate the usage of unlabeled target domain speech to enhance MDD performance. 

\noindent \textbf{Pseudo-labeling in speech domain.} PL has been a widely used approach for semi-supervised ASR. In general, these methods can be divided into offline and online PL by the generation scheme of pseudo labels. Offline PL methods use a separately-trained teacher model to assign pseudo labels for unlabeled samples. A student model is then trained on labeled plus pseudo-labeled samples \cite{xu2021self}. Filtering heuristics \cite{kahn2020self, park20d_interspeech} and iterative training \cite{xu20b_interspeech} were shown to be useful to improve PL quality. On the other hand, in online PL methods, pseudo labels are generated \textit{on-the-fly} by the online model itself \cite{higuchi2021momentum, chen20m_interspeech, zhu2021wav2vec}. We adopt the PL method in \cite{higuchi2021momentum}, but unlike \cite{higuchi2021momentum}, we combine PL with Wav2vec 2.0 fine-tuning, with phonemes as targets.

\begin{figure}[t]
    \centering
    \includegraphics[width=0.95\columnwidth]{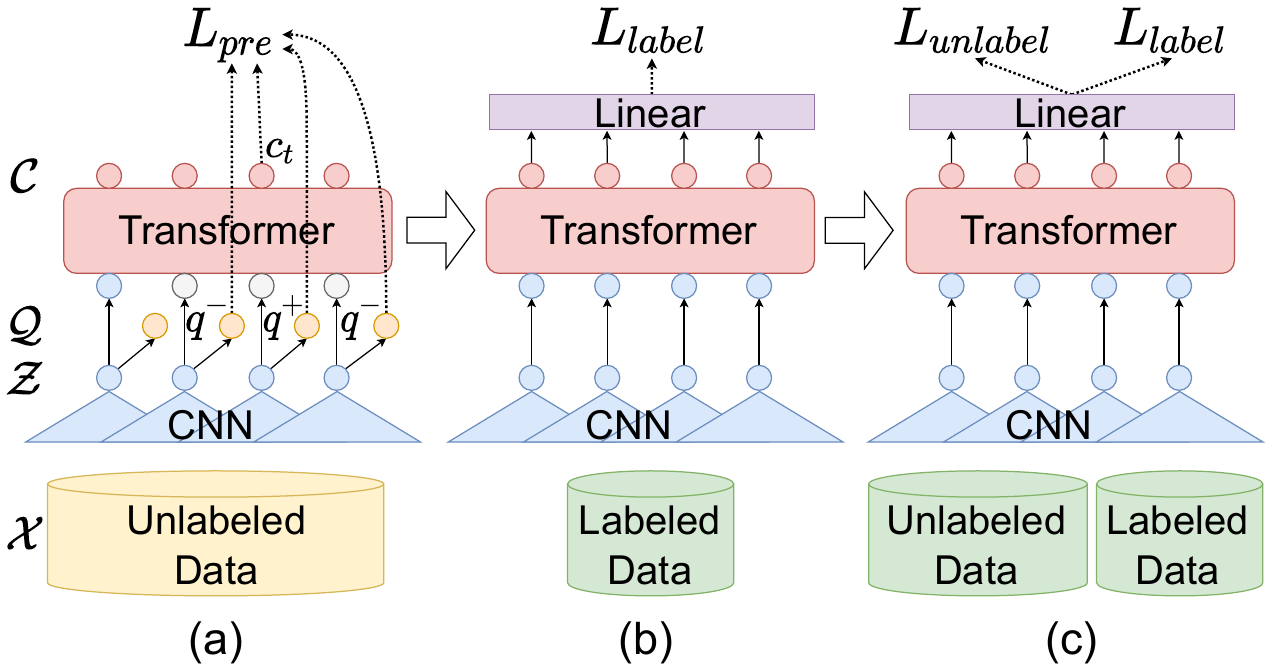}
    \caption{Three-stage training: we extend the conventional pre-training (a) + fine-tuning (b) pipeline with an additional fine-tuning step with MPL (c). Training data from different domains are marked with different colors. Better viewed in color.}
    \label{fig:system}
    \vspace{-4mm}
\end{figure}

\section{Method}

We first review the pre-training and fine-tuning of Wav2vec 2.0 (Section. \ref{sec:w2v2} and \ref{sec:ft}), and then describe our proposed fine-tuning method with MPL in Section. \ref{sec:ft_mpl}.

\subsection{Wav2vec 2.0} \label{sec:w2v2}

 Wav2vec 2.0 consists of Convolutional Neural Network (CNN) and Transformer layers (Figure. \ref{fig:system} (a)). The CNNs work as a feature extractor which converts input audio waveform $\mathcal{X}$ into latent representation $\mathcal{Z}$. Before fed to Transformer layers, $\mathcal{Z}$ is randomly masked by a certain portion (marked by grey in Figure. \ref{fig:system} (a)). The Transformer layers then contextualize $\mathcal{Z}$ into $\mathcal{C}$. The unmasked latent representation $\mathcal{Z}$ is further discretized to $\mathcal{Q}$ via a learnable codebook. Given the contextualized representation $c_{t}$ at masked time step $t$, we denote the discretized representation for time step $t$ as $q^+$, and $q^-$ for other masked steps. During pre-training, Wav2vec 2.0 is trained by Contrastive Loss which aims to distinguish the true underlying discretized representation $q^+$ for each masked step $t$ from those at other masked positions ($q^-$), based on the contextualized representation $c_{t}$. The full SSL loss, denoted as $\mathcal{L}_{pre}$, is a weighted sum of the Contrastive Loss and a codebook diversity loss \cite{baevski2020wav2vec}. Note that the model trained with such unsupervised SSL objective can be further fine-tuned by an ASR task using the text transcriptions of the pre-training audio, if available, to gain audio-text supervision \cite{baevski2020wav2vec}. We discuss the impact of this additional supervision on the downstream MDD task in Section. \ref{sec:960vs}.

\subsection{Fine-tuning} \label{sec:ft}

After pre-training, we add linear layers on top of the Transformer layers, and remove the discretization module. The entire model (with CNN layers frozen) is fine-tuned end-to-end on the downstream L2 speech phoneme recognition task using connectionist temporal classification (CTC) loss \cite{graves2006connectionist} (Figure. \ref{fig:system} (b)). Let $X=(x_1, x_2, ..., x_T)$ denote an input audio waveform, $Y=(y_1, y_2, ..., y_L)$ denote training targets, which in our MDD task are the human-labeled phoneme sequences (i.e. what the L2 speaker actually pronounced). The CTC loss $L_{label}$ on labeled samples can be expressed as

\vspace{-1mm}
\begin{equation}
    L_{label}(\theta)=-\log \sum_{A \in \beta(Y)} \prod_{i} P_{\theta}(a_{i}|X)
    \label{eq:ctc_label}
\end{equation}
where $A=(a_1, a_2 ... a_T)$ denotes a compatible latent alignment between $X$ and $Y$, $\beta(Y)$ denotes the set of all such compatible alignments, and $\theta$ denotes model parameters. We apply speed perturbation \cite{ko15_interspeech} on $X$, plus a modified SpecAugment on latent representations $\mathcal{Z}$ as in \cite{baevski2020wav2vec} for data augmentation.

\begin{table*}[t]
\caption{MDD evaluation metrics and PER on L2-ARCTIC test set. Numbers before and after ``/" represent the percentage and the absolute number of occurrences of a particular case. P and R stand for Precision and Recall, respectively. PER numbers with a star ($^{*}$) are not directly comparable to ours due to different settings (see Section. \ref{res:vsprior}).} \label{res:main}
\resizebox{1.0\textwidth}{!}{
\begin{tabular}{cl|cc|ccc|ccc|c}
\hline
\multicolumn{2}{c|}{\multirow{3}{*}{Models}} &
  \multicolumn{2}{c|}{Correct Pronunciations} &
  \multicolumn{3}{c|}{Mispronunciations} &
  \multirow{3}{*}{P (\%)} &
  \multirow{3}{*}{R (\%)} &
  \multirow{3}{*}{F1 (\%)} &
  \multirow{3}{*}{PER (\%)} \\ \cline{3-7}
\multicolumn{2}{c|}{} &
  \multirow{2}{*}{\begin{tabular}[c]{@{}c@{}}True Accept\\ (\%/\#)\end{tabular}} &
  \multirow{2}{*}{\begin{tabular}[c]{@{}c@{}}False Reject\\ (\%/\#)\end{tabular}} &
  \multirow{2}{*}{\begin{tabular}[c]{@{}c@{}}False Accept\\ (\%/\#)\end{tabular}} &
  \multicolumn{2}{c|}{True Reject (\%/\#)} &
   &
   &
   &
   \\ \cline{6-7}
\multicolumn{2}{c|}{}                                                   &             &           &            & Corr. Diag. & Err. Diag. &       &       &       &       \\ \hline
\multicolumn{1}{c|}{\multirow{3}{*}{Prior}} & CTC-Attn + Anti-Phone \cite{yan20_interspeech}     & --          & --        & --         & --          & --         & 46.57 & 70.28 & 56.02 & --    \\
\multicolumn{1}{c|}{}                       & wav2vec2-large-lv60 \cite{peng21e_interspeech}      & 94.01/24198 & 5.99/1542 & 43.37/1850 & 68.08/1645  & 31.91/771  & 61.04 & 56.63 & 58.75 & 16.01$^{*}$ \\
\multicolumn{1}{c|}{}                       & wav2vec2-large-XLSR \cite{peng21e_interspeech}      & 94.57/24343 & 5.43/1397 & 43.95/1875 & 65.75/1572  & 34.25/819  & 63.12 & 56.05 & \textbf{59.37} & 15.43$^{*}$ \\ \hline
\multicolumn{1}{c|}{\multirow{6}{*}{Ours}}  & wav2vec2-base             & 92.84/23873 & 7.16/1841 & 46.07/1977 & 75.84/1755  & 24.16/559  & 55.69 & 53.93 & 54.80 & 15.52 \\
\multicolumn{1}{c|}{}                       & \quad + one-shot PL (scratch)   & 93.26/23982 & 6.74/1732 & 46.05/1976 & 76.11/1762  & 23.89/553  & 57.20 & 53.95 & 55.53 & 14.85 \\
\multicolumn{1}{c|}{}                       & \quad + one-shot PL (continual) & 93.16/23955 & 6.84/1759 & 46.17/1981 & 76.19/1760  & 23.81/550  & 56.77 & 53.83 & 55.26 & 15.04 \\
\multicolumn{1}{c|}{}                       & \quad + MPL                     & 93.54/24052 & 6.46/1662 & 45.84/1967 & \textbf{77.24/1795}  & \textbf{22.76/529}  & 58.30 & 54.16 & \textbf{56.16} & \textbf{14.69} \\ \cline{2-11} 
\multicolumn{1}{c|}{}                       & wav2vec2-base-960h        & 93.83/24128 & 6.17/1586 & 47.91/2056 & 76.64/1713  & 23.36/522  & 58.49 & 52.09 & 55.10 & 14.87 \\
\multicolumn{1}{c|}{}                       & \quad + MPL                     & 94.40/24273 & 5.60/1441 & 48.80/2094 & \textbf{77.29/1698}  & \textbf{22.71/499}  & 60.39 & 51.20 & \textbf{55.42} & \textbf{14.36} \\ \hline
\end{tabular}
}
\end{table*}

\subsection{Fine-tuning with momentum pseudo-labeling} \label{sec:ft_mpl}

In addition to the fine-tuning stage where only labeled samples are used, we consider including unlabeled samples from the target domain into fine-tuning (Figure. \ref{fig:system} (c)). Formally, given the available labeled L2 speech samples $\mathcal{D}^L$ (same samples as in Figure. \ref{fig:system} (b)) and additional unlabeled speech samples $\mathcal{D}^U$, our goal is to continually learn a new model $\xi$ based on a base model $\theta$ learned in the previous fine-tuning stage, using both $\mathcal{D}^L$ and $\mathcal{D}^U$. We propose to leverage the unlabeled samples via momentum pseudo-labeling (MPL) \cite{higuchi2021momentum}. In MPL, an offline teacher model $\phi$ is used to assign pseudo labels for the unlabeled samples, which guide the learning of an online student model $\xi$. Pseudo labels $\hat{Y}$ are inferred by the teacher model $\phi$:

\vspace{-1mm}
\begin{equation}
    \hat{Y}=\underset{Y}{\operatorname{argmax}} \; P_{\phi}(Y \mid X), \; X \in \mathcal{D}^{U}
\end{equation}
where we use $\operatorname{argmax}$ to represent greedy CTC decoding. Then, similar to Equation. \ref{eq:ctc_label}, $\hat{Y}$ can be used as the training targets of the unlabeled samples for the student model $\xi$:

\vspace{-3mm}
\begin{equation}
L_{unlabel}(\xi)=-\log \sum\limits_{A \in \beta(\hat{Y})} \prod\limits_{i} P_{\xi}(a_{i}|X), \; X \in \mathcal{D}^{U}
\end{equation}
The online model can then be trained on both labeled and unlabeled samples using a unified loss $L$:

\begin{equation}
    L(\xi) = \left\{\begin{aligned}
        & L_{label}(\xi)  & (X, Y) &\in \mathcal{D}^{L} \\ 
        & L_{unlabel}(\xi) &  X &\in \mathcal{D}^{U}
    \end{aligned}\right.
    \label{eq:unified_loss}
\end{equation}
To create dynamic pseudo labels, the teacher model $\phi$ is also updated during training by a moving average of itself and the online student model $\xi$, controlled by a momentum factor $\alpha$:
\begin{equation}
    \phi \leftarrow \alpha \phi + (1-\alpha) \xi
    \label{eq:momentum}
\end{equation}
The momentum factor here controls the update magnitude of $\phi$ and makes $\phi$ evolve more smoothly than $\xi$, which prevents the pseudo labels from changing drastically and helps stabilize training. We follow the heuristic in \cite{higuchi2021momentum} to determine $\alpha$ based on total training steps. Both $\xi$ and $\phi$ are initialized with $\theta$ before entering MPL fine-tuning, and are updated according to Equation. \ref{eq:unified_loss} and \ref{eq:momentum} respectively at each training step. The data augmentation in Section. \ref{sec:ft} is also applied in this stage. Compared to static PL methods \cite{kahn2020self, park20d_interspeech}, since the teacher model can be seen as an ensemble of the online model at different training steps, MPL is expected to be more robust to pseudo label noise. We compare MPL with static PL in Section. \ref{res:mpl}.

\section{Experimental setup}
\subsection{Datasets} \label{sec:data}

\textbf{L2-ARCTIC} \cite{zhao18b_interspeech} is used to train our MDD model. L2-ARCTIC includes L2 speech from 24 non-native English speakers with different L1 backgrounds (Indian, Mandarin, Vietnamese, Korean, Arabic, Spanish). It provides human-labeled perceived phonemes for around 15\% of the utterances per speaker. Following \cite{peng21e_interspeech, feng2020sed}, we set labeled samples from 6 speakers as test set and labeled samples for the remaining speakers as labeled training set. Unlabeled samples of the remaining 18 speakers are used as the unlabeled training set. We randomly split 10\% of the labeled training set as development set. Statistics of our data splits are shown in Table. \ref{tab:data}. Since L2-ARCTIC uses artificial \textit{sil} tokens to represent phoneme deletions and insertions, to construct target training phoneme sequences, we remove the artificial \textit{sil} tokens, while preserving the \textit{sil} tokens that correspond to true pauses and silences. 

\noindent\textbf{UTD-4Accents} \cite{ghorbani18_interspeech} is an in-house dataset that consists of 4 English accents: US (native), Australian, Spanish and Indian. We use the Indian-accent part for the open test (see Section. \ref{res:open_test}), which includes 112 speakers (balanced for gender and age). The utterances are read speech from diverse domains, including general vocabulary, voice search, etc.

\begin{table}[b]
    \centering
    \vspace{-2mm}
    \caption{L2-ARCTIC data splits and statistics.} \label{tab:data}
    \vspace{-1mm}
    \begin{tabular}{l|cc|c|c}
    \hline
                  & \multicolumn{2}{c|}{Train} & \multirow{2}{*}{Development} & \multirow{2}{*}{Test} \\ \cline{2-3}
                  & labeled     & unlabeled    &                      &                       \\ \hline
    \# utterances & 2429        & 17381        & 268                  & 900                   \\
    \# hours      & 2.51        & 17.73        & 0.27                 & 0.88                  \\ \hline
    \end{tabular}
    
\end{table}

\subsection{Evaluation}

We evaluate the PER between recognized phonemes and training targets, as well as the MDD metrics following \cite{peng21e_interspeech, li2016mispronunciation}. For cases of correct pronunciations where human-perceived phonemes agree with canonical phonemes (i.e. the phonemes that a L2 speaker was supposed to pronounce), we have True Accept (TA) and False Reject (FR) cases based on whether model predictions match both canonical and perceived phonemes. In the same spirit, for mispronunciation cases where human perceived phonemes are inconsistent with canonical phonemes, we could have False Accept (FA) and True Reject (TR) cases. TR can be further divided into Correct Diagnosis and Erroneous Diagnosis, based on whether model predictions match human labels. Precision (P), Recall (R) and F1 can then be computed from TR, FR, FA: $P = TR/(FR+TR); R = TR/(FA+TR); F1=2PR/(P+R)$.

\subsection{Implementation details}

We tune hyper-parameters on the development set. The best model (in terms of PER) on the development set is evaluated on the test set. We experimented with two pre-trained HuggingFace \cite{wolf2020transformers} Wav2vec 2.0 models, \textit{wav2vec2-base} and \textit{wav2vec2-base-960h}. Both models have identical base-size network architecture and are pre-trained with the same SSL objective (Section. \ref{sec:w2v2}) on 960-hour LibriSpeech audio \cite{panayotov2015librispeech}. \textit{wav2vec2-base-960h} has been additionally fine-tuned by an ASR task on LirbiSpeech after pre-training, which learns explicit audio-text mapping compared to \textit{wav2vec2-base}. We use separate Adam optimizers for linear layers and Wav2vec 2.0 layers, with fixed learning rates of $3e-4$ and $1e-5$, respectively. Gradient accumulation is used to obtain an effective batch size of 32. All experiments run for 50 epochs on a single NVIDIA RTX 2080 GPU. Our implementation is based on SpeechBrain toolkit \cite{speechbrain}. 

\section{Results}

\subsection{Effect of pseudo-labeling and pre-trained models} \label{sec:960vs}

In Table. \ref{res:main}, we use \textit{wav2vec2-base} and \textit{wav2vec2-base-960h} to denote the vanilla fine-tuning baselines based on the corresponding pre-trained SSL models without using any unlabeled samples (i.e. stage (b) in Figure. \ref{fig:system}). First, we can see that with MPL, both \textit{wav2vec2-base} and \textit{wav2vec2-base-960h} gain a significant improvement over the vanilla fine-tuning baseline, in terms of F1 score (\textit{wav2vec2-base}: 56.16 vs. 54.80; \textit{wav2vec2-base-960h}: 55.42 vs. 55.10) and PER (\textit{wav2vec2-base}: 14.69 vs. 15.52; \textit{wav2vec2-base-960h}: 14.36 vs. 14.87). This demonstrates the benefit of leveraging unlabeled L2 samples. Second, an interesting finding is that \textit{wav2vec2-base-960h} produces more False Accepts, less False Rejects and less overall Reject cases than \textit{wav2vec2-base}, leading to higher Precision but lower Recall. This means that it tends to be a more ``tolerant" judge by rejecting less L2 pronunciations. One possible reason is that the extra audio-text supervision has biased it to the canonical pronunciations, which makes it ``over-robust" to mispronunciations. However, for MDD task this may not be desired, because we expect the MDD model to faithfully reflect what a L2 speaker actually pronounced and the ``over-robustness" may conceal the mispronunciations. In contrast, \textit{wav2vec2-base} has a more balanced Precision and Recall. 

\begin{figure}[t]
    \centering
    \includegraphics[width=0.95\columnwidth]{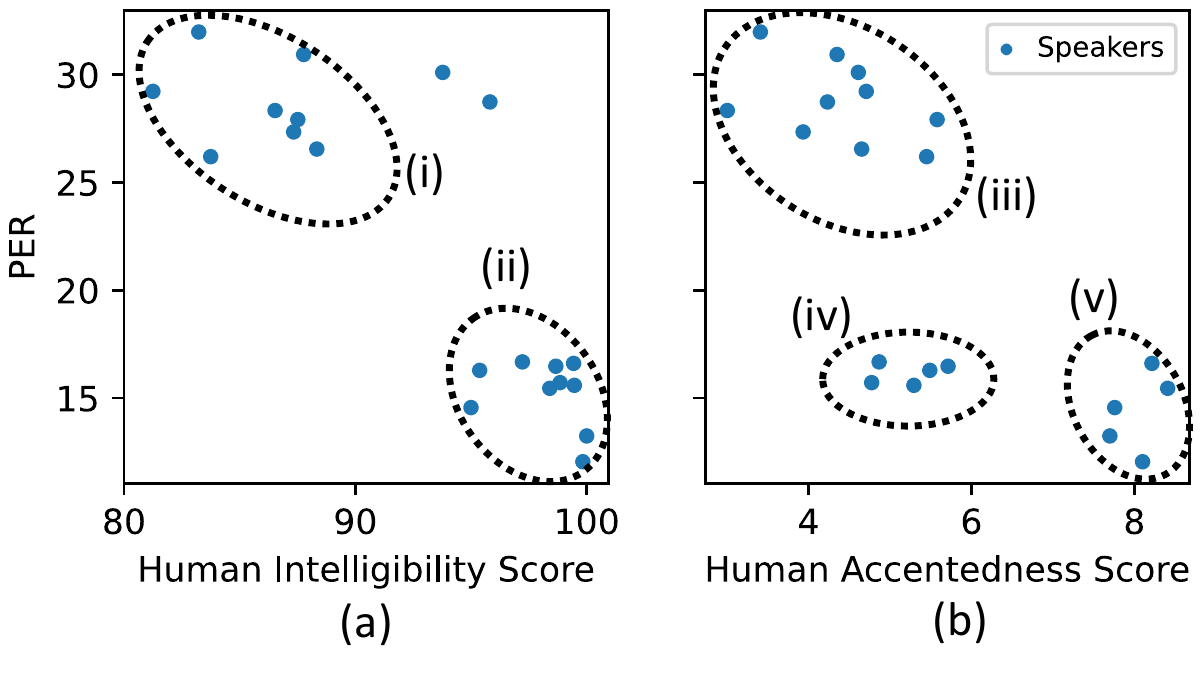}
    \vspace{-3mm}
    \caption{(a) PER vs. Human Intelligibility Score. Pearson Correlation $=-0.84$ ($p<10^{-5}$); (b) PER vs. Human Accentedness Score. Pearson Correlation $=-0.75$ ($p<10^{-3}$).}
    \label{fig:corr}
    \vspace{-4mm}
\end{figure}

\subsection{Effect of the momentum update mechanism} \label{res:mpl}

To study whether MPL is superior to static PL methods, we compare MPL with two PL baselines: \textbf{(1) one-shot PL (scratch)}: pseudo labels are generated offline by the model fine-tuned with labeled samples only (i.e. stage (b) in Figure. \ref{fig:system}). A new model is then trained from scratch using the pseudo-labeled samples plus the original labeled samples; \textbf{(2) one-shot PL (continual)}: same as the above method, except that the new model is initialized with the weights of the fine-tuned model. Since in both baselines pseudo labels are fixed once generated, we refer to them as \textit{one-shot}. Table. \ref{res:main} compares MPL with the two baselines on \textit{wav2vec2-base}. Although both baselines outperform the labeled-samples-only fine-tuning, a larger improvement comes from MPL, demonstrating the benefits of having dynamic pseudo labels. The performances of the two baselines are comparable, with \textit{one-shot PL (continual)} being slightly worse. This may be caused by over-fitting.

\subsection{Comparison with prior works} \label{res:vsprior}

Finally, we compare our MDD models with the current leading MDD methods. \cite{yan20_interspeech} uses a CTC-Attention model with Anti-Phone augmentation. \cite{peng21e_interspeech} is also based on fine-tuning Wav2vec 2.0 models and achieves state-of-the-art MDD performance. They used large-size Wav2vec 2.0 models which are pre-trained on larger-scale audio corpora. Models in \cite{peng21e_interspeech} have 300M+ parameters, while ours have around 90M parameters.\footnote{Due to computation resources limit, we are not able to run experiments on large models, but our proposed methods are generic and we leave applications on larger-size models for future investigation.} From Table. \ref{res:main}, we can see that \cite{yan20_interspeech} achieves higher Recall but much lower Precision. Our proposed MPL model outperforms \cite{yan20_interspeech} in terms of overall F1. Compared with \cite{peng21e_interspeech}, although our proposed method does not outperform theirs, we observe a higher Correct Diagnosis and lower Erroneous Diagnosis in the percentage and the absolute number of occurrences. This implies that our models are able to provide more accurate pronunciation diagnosis feedback which may help L2 learners correct their mispronunciations more effectively. Note that the PER reported in \cite{peng21e_interspeech} is not directly comparable to ours, as they did not pre-process the target phoneme sequences as we do (Section. \ref{sec:data}).

\subsection{Open test: Indian accent and intelligibility assessment} \label{res:open_test}
 We hypothesize that a proper MDD model should perceive the L2 pronunciations in a similar way as humans, and thus there should exist a correlation between its phoneme recognition performance and L2 speech accentedness and comprehensibility, i.e. a higher PER (more mispronunciations) corresponds to a heavier accent and lower comprehensibility. Our goal is to investigate the existence of such relations, which shed some light on the applicability of the MDD model towards automatic intelligibility assessment for L2 speech. 

For this purpose, we run an open test of our best MDD model (\textit{wav2vec2-base} + MPL) on Indian-accented L2 speech from the UTD-4Accents dataset. We compute per-speaker PERs between the recognized phonemes and the canonical phonemes given by a grapheme-to-phoneme model.\footnote{\url{https://github.com/Kyubyong/g2p}} Then we select 10 highest-PER speakers and 10 lowest-PER speakers, and randomly sample 10 utterances for each of the 20 speakers. 17 human listeners score the accentedness (scale: 1-9 where 1 means heavy accent) and intelligibility (scale: 0-100 where 0 means not intelligible at all) of the sampled utterances. Each speaker receives 30+ ratings from different raters. We then aggregate per-speaker accentedness and intelligibility scores. All raters are graduate students at Northern Arizona University with 2+ years of experience in L2 pronunciation teaching or research. To create unbiased ratings, raters are presented with the L2 audio only, without any text transcription or other information. 

We plot the per-speaker PER-Intelligibility and PER-Accentedness relations in Figure. \ref{fig:corr} (a) and (b), respectively. Figure. \ref{fig:corr} (a) shows that the speakers are roughly grouped into 2 clusters ((i) and (ii)). This is consistent with our expectation: since we selected 20 speakers with highest and lowest PER, we expect those speakers are also clustered into more-intelligible and less-intelligible groups. Further, the PERs and human intelligibility scores present a strong negative correlation, which means that higher PERs align with less intelligible speech. Such a negative correlation can also be observed in the PER-Accentedness plot (Figure. \ref{fig:corr} (b)). Interestingly, besides the two aforementioned clusters, we observe an additional cluster (iv). Speakers in this cluster reside in cluster (ii) in  Figure. \ref{fig:corr} (a). This indicates that for these speakers, humans perceive relatively heavy accents, while they are still highly intelligible. This is possibly because apart from phonetic errors, accentedness is also highly impacted by prosodic factors, such as intonation, lexical stress, etc, which may not be as important for intelligibility. Since our MDD model only detects phonetic-level errors, assessing the accentedness of these speaker is beyond its capacity. In summary, the alignment between phoneme recognition of the MDD model and human perception has validated our motivation of using the MDD model as a component towards automatic L2 speech intelligibility assessment. 

\section{Conclusions and acknowledgements}

We have presented an approach to use unlabeled L2 speech to enhance MDD performance via pseudo-labeling. In addition, we take one step forward towards using the MDD model for automatic L2 speech intelligibility and accentedness assessment. Through a human listening test, we have shown that the MDD model recognition performance shows a strong correlation with human perception. In future, we plan to include more speech attributes, such as lexical stress, speech rate, into the L2 speech intelligibility assessment framework. This study is supported by NSF EAGER CISE Project 2140415, and partially by the University of Texas at
Dallas from the Distinguished University Chair in Telecommunications Engineering held by J. H. L. Hansen. We would like to thank the raters at Northern Arizona University for their participation in our listening test.



\pagebreak
\bibliographystyle{IEEEtran}

\bibliography{mybib}

\end{document}